\documentclass[aps,preprint,a4paper,12pt,onecolumn]{revtex4}%
\usepackage{amsfonts}
\usepackage{amsmath}
\usepackage{amssymb}
\usepackage{graphicx}%
\providecommand{\U}[1]{\protect\rule{.1in}{.1in}}

\begin{document}
\title{Theoretical method for the generation of a dark two-mode squeezed state of a
trapped ion}
\author{T. Werlang and C. J. Villas-Boas}
\affiliation{Departamento de F\'{\i}sica, Universidade Federal de S\~{a}o Carlos, Caixa
Postal 676, 13595-905, S\~{a}o Carlos, S\~{a}o Paulo, Brasil}

\begin{abstract}
Here we show how to generate a dark two-mode squeezed state of a trapped ion,
employing a three-level ion in a V configuration with a strong decay of the
excited states. The degree of squeezing can be manipulated by choosing the
intensity of the driving fields. Our scheme is robust against the usual
dissipation mechanism and could be implemented with present-day technology.
The validity of the approximations employed in this work was tested by
numerical calculations, which agreed completely with the analytical solutions.

\end{abstract}
\maketitle

The recent experimental advances in Quantum Optics, specially in the domain of
trapped ions, have allowed fundamental features of quantum mechanics, such as
geometric phases \cite{GPhase-ions} and Bell inequalities \cite{Bell-ions}, to
be investigated, as well as offering potential applications in quantum
computation \cite{nielsen-chuang, quantum-computation-ion} and teleportation
processes \cite{teleport-2004}. With the advent of quantum information theory,
the generation of entangled states has became essential for the implementation
of quantum communication protocols \cite{nielsen-chuang} and to improve our
understanding of this non-local character of the quantum theory
\cite{entanglement}. In particular, the two-mode vacuum squeezed (TMVS) state,
i.e., the original Einstein-Podolsky-Rosen (EPR) state \cite{epr-1935}, has
attracted much attention because it can show a high degree of entanglement
\cite{entanglement-gaussian-state} and can be useful for teleportation of
continuous variable states \cite{brausntein-teleport}. Success in generating
the TMVS state has been reported in the running wave domain, with a parametric
down conversion process \cite{brausntein-teleport}. However, the experimental
generation of this state in the cavity quantum electrodynamics (QED) or
trapped ion domains has not been achieved so far, mainly due to the
sensitivity of quantum states to system-environment interaction. In the cavity
QED context, several theoretical schemes with three-level atoms
\cite{serra-pra2005, celso-epjd2005, solano-prl2006}, or even two-level atoms,
where the sideband transition is used \cite{prado-pra2006}, have been
elaborated for the generation of the TMVS state. Also in the trapped ion
domain we find some schemes which allow the generation of this entangled state
through the manipulation of laser fields \cite{two-mode-ideal-ion}. However,
none of the schemes cited above take into account the system-environment
interaction, which degrades the quantum states so that, in general, the
fidelity of the generated states decays quickly. In this scenario,
\textit{reservoir engineering} appears to offer a possible way round this
problem and can generate robust non-classical states of the radiation field or
of the ionic motion. For example, using the atomic decay of the internal
levels of a single ion, in Ref. \cite{cirac-prl1993} the authors showed how to
construct a reservoir able to lead the motion of the ion to an squeezed state
asymptotically. Similar schemes have been employed to protect various
superpositions of coherent states \cite{matos-prl1996, matos-pra1996,
gou-pra1997}, and in ref. \cite{carvalho-prl2001} the authors showed how to
protect any one-dimensional motional state of an ion against decoherence. Also
in this context of reservoir engineering, in Ref. \cite{tombesi} we find an
effective master equation that, in the stationary state, filters specific
number states of the vibrational motion of a trapped ion. On the other hand,
reservoir engineering for multi-mode states has been addressed only recently,
and there are few theoretical schemes so far. We can cite, for example, in the
trapped ion domain, theoretical schemes for the preparation of a pair coherent
state \cite{gou-pra1996a} and pair cat states \cite{gou-pra1996b}, SU(1,1)
intelligent states \cite{gerry-pra1997} and dark SU(2) states of a trapped ion
\cite{kis-pra2001}. Recently, Parkins \textit{et al.} \cite{parkins-prl2006}
has proposed a scheme for the unconditional generation of a two-mode squeezed
state of two separated atomic ensembles. A similar scheme was employed for the
generation of the TMVS state for the motion of two ions in different traps
\cite{li-pra2006} or even a single ion in a two-dimension trap
\cite{chines-referee} inside an optical cavity. In Ref.
\cite{davidovich-prl2007} it was shown theoretically how to generate this
entangled state using an atomic reservoir for a two-mode cavity. In Ref.
\cite{gou1996}, the authors showed how a beam splitter operation may be
produced in a single ion in two-dimension trap. After generating a robust
squeezed state of a single mode of a trapped ion \cite{cirac-prl1993}, this
effective interaction could be directly employed to generate a TMVS state, but
in this case the TMVS state would not be the steady state of the system.

In this communication we report a simple feasible scheme for the unconditional
generation of the TMVS state in a single trapped ion. For this purpose we have
employed a two-dimensional harmonic motion (on the $x$ and $y$ axes) of the
center of mass of a single ion in a $V$ configuration (see Fig. 1). The
excited states, $\left\vert 1\right\rangle $ and $\left\vert 2\right\rangle $,
are coupled to the ground state $\left\vert 0\right\rangle $ through classical
fields (propagating along the $x$ and $y$ axes). When the decay of the excited
electronic states is stronger than the effective coupling between the
vibrational and the internal ionic states, the steady state of the vibrational
modes, for convenient choices of the intensity and frequency of the classical
fields, turns out to be exactly the TMVS state. Even in the presence of a
thermal reservoir for the ionic motion, we show that the generated TMVS state
is almost exactly the desired one. Such a scheme, based on reservoir
engineering, is robust against dissipative effects of the vibrational modes
and could be used to investigate experimentally the entanglement properties of
this state. The basic level configuration needed for the implementation of our
scheme is sketched in Fig. 1. We consider an ion with mass $m$ in a
two-dimensional trap, driven by four classical fields, two of them along
the$\ x$ axis and the other two along the $y$ axis, with complex amplitudes
$\Omega_{j\alpha}=\left\vert \Omega_{j\alpha}\right\vert e^{i\varphi_{j\alpha
}}$ ($\left\vert \Omega_{j\alpha}\right\vert $ being the Rabi frequency and
$\varphi_{j\alpha}$ the phase of the classical fields), frequencies
$\omega_{j\alpha}$, and wave numbers $k_{i\alpha}$, $j=1,2$ and $\alpha=x,y$.
The total Hamiltonian of the system is given by $H=H_{0}+V(t)$, with
\begin{subequations}
\label{eq1}%
\begin{align}
H_{0} &  =\hbar\omega_{1}\sigma_{11}+\hbar\omega_{2}\sigma_{22}+\hbar\nu
_{x}a^{\dagger}a+\hbar\nu_{y}b^{\dagger}b,\label{eq1a}\\
V(t) &  =\hbar\left[  \Omega_{1x}e^{ik_{1x}x-i\omega_{1x}t}+\Omega
_{1y}e^{ik_{1y}y-i\omega_{1y}t}\right]  \sigma_{10}\nonumber\\
&  +\hbar\left[  \Omega_{2x}e^{ik_{2x}x-i\omega_{2x}t}+\Omega_{2y}%
e^{ik_{2y}y-i\omega_{2y}t}\right]  \sigma_{20}+h.c.,\label{eq1b}%
\end{align}
where $\omega_{1}$ and $\omega_{2}$ stand for the atomic transition
frequencies between the states $\left\vert 0\right\rangle \longleftrightarrow
\left\vert 1\right\rangle $ and $\left\vert 0\right\rangle \longleftrightarrow
\left\vert 2\right\rangle $ respectively, $\sigma_{lm}=\left\vert
l\right\rangle \left\langle m\right\vert $, $l,m=0,1,2$, are the atomic
operators, $\ a$ ($b$) and $a^{\dagger}$ ($b^{\dagger}$) are the annihilation
and creation operators of the vibrational mode in the $x$ ($y$) axis, with
frequency $\nu_{x}$ ($\nu_{y}$), and $h.c.$ stands for Hermitian conjugate. In
the Lamb-Dicke limit, i. e., $\eta_{j\alpha}\ll1$, where $\eta_{j\alpha
}=k_{j\alpha}\sqrt{\frac{\hbar}{2m\nu_{ja}}}$ , $j=1,2$ and $\alpha=x,y$, the
above Hamiltonian can be written in the interaction picture as
\end{subequations}
\begin{align}
H_{I} &  =\hbar\Omega_{1x}e^{-i\delta_{1x}t}\left[  1+i\eta_{1x}\left(
ae^{-i\nu_{x}t}+a^{\dagger}e^{i\nu_{x}t}\right)  \right]  \sigma
_{10}\nonumber\\
&  +\hbar\Omega_{1y}e^{-i\delta_{1y}t}\left[  1+i\eta_{1y}\left(
be^{-i\nu_{y}t}+b^{\dagger}e^{i\nu_{y}t}\right)  \right]  \sigma
_{10}\nonumber\\
&  +\hbar\Omega_{2x}e^{-i\delta_{2x}t}\left[  1+i\eta_{2x}\left(
ae^{-i\nu_{x}t}+a^{\dagger}e^{i\nu_{x}t}\right)  \right]  \sigma
_{20}\nonumber\\
&  +\Omega_{2y}e^{-i\delta_{2y}t}\left[  1+i\eta_{2y}\left(  be^{-i\nu_{y}%
t}+b^{\dagger}e^{i\nu_{y}t}\right)  \right]  \sigma_{20}+h.c.,\label{eq2}%
\end{align}
with $\delta_{i\alpha}=\omega_{i\alpha}-\omega_{i}$. Supposing $\delta
_{1x}=-\delta_{2x}=-\nu_{x}$, $\delta_{1y}=-\delta_{2y}=\nu_{y}$, $\left\vert
\delta_{i\alpha}\right\vert \gg\left\vert \eta_{i\alpha}\Omega_{i\alpha
}\right\vert $, and applying a rotating-wave approximation, the effective
Hamiltonian becomes
\begin{equation}
H_{I}=\hbar\lambda_{1x}\left\{  a+\frac{\lambda_{1y}}{\lambda_{1x}}b^{\dagger
}\right\}  \sigma_{10}+\hbar\lambda_{2y}\left\{  b+\frac{\lambda_{2x}}%
{\lambda_{2y}}a^{\dagger}\right\}  \sigma_{20}+h.c.\label{eq3}%
\end{equation}
where we have defined $\lambda_{j\alpha}\equiv i\eta_{j\alpha}\Omega_{i\alpha
}$. As in Ref. \cite{parkins-prl2006}, we can apply a unitary transformation
$\tilde{\rho}=S_{ab}^{\dagger}(\xi)\rho S_{ab}(\xi)$, with $S_{ab}(\xi
)=\exp\left(  \xi^{\ast}ab-\xi a^{\dagger}b^{\dagger}\right)  $ and
$\xi=e^{i\phi}r$, this last being the two-mode squeezing operator ($r$ stands
for the squeezing factor and $\phi$ the angle of squeezing), to obtain the
transformed Hamiltonian
\begin{equation}
\widetilde{H}_{I}=S_{ab}^{\dagger}(\xi)H_{I}S_{ab}(\xi)=\hbar\widetilde
{\lambda}_{a}a\sigma_{10}+\hbar\widetilde{\lambda}_{b}b\sigma_{20}%
+h.c.,\label{eq4}%
\end{equation}
where $\widetilde{\lambda}_{a}=\lambda_{1x}\cosh(r)-e^{-i\phi}\lambda
_{1y}\sinh(r)$, $\widetilde{\lambda}_{b}=\lambda_{2y}\cosh(r)-e^{-i\phi
}\lambda_{2x}\sinh(r),$ and we have assumed $\lambda_{1y}\cosh(r)-e^{i\phi
}\lambda_{1x}\sinh(r)=\lambda_{2x}\cosh(r)-e^{i\phi}\lambda_{2y}\sinh(r)=0$,
which implies that\textbf{ }%
\begin{equation}
r=\operatorname{arctanh}\left\vert \frac{\lambda_{1y}}{\lambda_{1x}%
}\right\vert =\operatorname{arctanh}\left\vert \frac{\lambda_{2x}}%
{\lambda_{2y}}\right\vert \label{eqq4}%
\end{equation}
and
\begin{equation}
\phi=\varphi_{1x}-\varphi_{1y}=-\left(  \varphi_{2x}-\varphi_{2y}\right)
.\label{eqq4b}%
\end{equation}
In this way, the squeezing factor $r$ and the squeezing angle $\phi$ can be
manipulated, respectively, by the intensities and the phases $\varphi
_{j\alpha}$ of the classical fields. When we take into account the atomic
decay of levels $\left\vert 1\right\rangle $ and $\left\vert 2\right\rangle $,
decay rates $\Gamma_{1}$ and $\Gamma_{2}$ respectively, the dynamics of the
system, in the transformed picture, is determined by the master equation
\begin{equation}
\overset{\cdot}{\widetilde{\rho}}=-\frac{i}{\hbar}\left[  \widetilde{H}%
_{I},\widetilde{\rho}\right]  +\mathcal{L}_{1}\widetilde{\rho}+\mathcal{L}%
_{2}\widetilde{\rho},\label{eq5}%
\end{equation}
where $\mathcal{L}_{1}\widetilde{\rho}=\frac{\Gamma_{1}}{2}\left(
2\sigma_{01}\widetilde{\rho}\sigma_{10}-\sigma_{11}\widetilde{\rho}%
-\widetilde{\rho}\sigma_{11}\right)  $\ and $\mathcal{L}_{2}\widetilde{\rho
}=\frac{\Gamma_{2}}{2}\left(  2\sigma_{02}\widetilde{\rho}\sigma_{20}%
-\sigma_{22}\widetilde{\rho}-\widetilde{\rho}\sigma_{22}\right)  $. The steady
state of Eq. (\ref{eq5}) is the vacuum for both modes and $\left\vert
0\right\rangle $ for the electronic state. We have assumed a strong decay of
both excited electronic states once, as pointed in Ref. \cite{cirac-prl1993},
we need two distinct dissipation channels ($\mathcal{L}_{1}\widetilde{\rho}$
and $\mathcal{L}_{2}\widetilde{\rho}$) to protect a two-mode quantum state of
a trapped ion. (Without this assumption we can not ensure that the steady
state of both modes, in the transformed picture, is the vacuum state.) By
applying the reverse unitary transformation, it is readily shown that the
steady state of this system is
\begin{equation}
\rho\left(  t\rightarrow\infty\right)  =S_{ab}(\xi)\tilde{\rho}S_{ab}%
^{\dagger}(\xi)=S_{ab}(\xi)\left\vert 0,0\right\rangle \left\langle
0,0\right\vert S_{ab}^{\dagger}(\xi)\otimes\left\vert 0\right\rangle
\left\langle 0\right\vert ,\label{asyntoptic}%
\end{equation}
which is a pure state for the vibrational modes $a$ and $b$, i.e., exactly the
two-mode squeezed vacuum state $\left\vert \Psi\right\rangle =\Sigma_{n}%
\tanh^{n}\left(  r\right)  /\cosh\left(  r\right)  \left\vert n,n\right\rangle
_{ab}$. The degree of squeezing $r$ is determined by the amplitudes of the
classical fields $\Omega_{j\alpha}$ since $\tanh\left(  r\right)  =\left\vert
\frac{\lambda_{1y}}{\lambda_{1x}}\right\vert =\left\vert \frac{\lambda_{2x}%
}{\lambda_{2y}}\right\vert $ and $\lambda_{j\alpha}\equiv i\eta_{j\alpha
}\Omega_{j\alpha}$. This steady state does not depend on the initial
electronic or motional state of the ion. Thus, the ion does not nave to be
cooled to the fundamental state in order to prepare such a state. Also, as the
entangled state is generated through the engineered reservoir, there is no
requirement for a precisely timed interaction between the ion and the laser
fields and the degree of entanglement ($r$) is determined only by the ratio of
the amplitudes of the classical fields (see Eq. (\ref{eqq4})). In this scheme,
the TMVS state is generated when the system reaches the steady state. As
pointed out in Refs. \cite{parkins-prl2006, li-pra2006}, the time needed for
the system to reach the steady state is defined by the atomic decay rate
$\Gamma$. For $\left\vert \widetilde{\lambda}_{a}\right\vert \sim\left\vert
\widetilde{\lambda}_{b}\right\vert >\Gamma$, this time will be of the order of
a few times $1/\Gamma$. In Ref. \cite{chines-referee} Tang \textit{et al.}
showed how to generate the TMVS state in a single ion in a two-dimension trap
inside a non-ideal optical cavity. Differently of our scheme, where the
required dissipation channels are played by the decay of the excited
electronic levels, in Ref. \cite{chines-referee} the required dissipation
channel is played by the decay of the cavity mode.

To check our result we solve numerically the master equation for our system in
the interaction picture
\begin{equation}
\overset{\cdot}{\rho}=-\frac{i}{\hbar}\left[  H_{I},\rho\right]
+\mathcal{L}_{1}\rho+\mathcal{L}_{2}\rho+\mathcal{L}_{ab}\rho, \label{eq6}%
\end{equation}
where $H_{I}$ is given by Eq. (\ref{eq3}). In this equation we have introduced
the Liouvillian $\mathcal{L}_{ab}\rho$ which describes the action of the
thermal reservoir on both atomic motions, i. e.,
\begin{align}
\mathcal{L}_{ab}\rho &  =%
{\displaystyle\sum\limits_{\alpha=a,b}}
\left\{  \frac{\left(  \overline{n}_{th}+1\right)  \gamma_{\alpha}}{2}\left(
2\alpha\rho\alpha^{\dagger}-\alpha^{\dagger}\alpha\rho-\rho\alpha^{\dagger
}\alpha\right)  \right. \nonumber\\
&  \left.  +\frac{\overline{n}_{th}\gamma_{\alpha}}{2}\left(  2\alpha
^{\dagger}\rho\alpha-\alpha\alpha^{\dagger}\rho-\rho\alpha\alpha^{\dagger
}\right)  \right\}  , \label{eq7}%
\end{align}
$\overline{n}_{th}$ being the mean number of quanta of the thermal reservoir
and $\gamma_{a}$ ($\gamma_{b}$) the decay rate of the vibrational mode $x$
($y$). We start with the ion in the internal ground state $\left\vert
0\right\rangle $ and both modes in the thermal state, $\rho_{ab}(0)=\rho
_{a}\otimes\rho_{b}$, with $\rho_{a}=\rho_{b}=\Sigma_{n=0}^{\infty}%
\frac{\overline{n}^{n}}{\left(  1+\overline{n}\right)  ^{n+1}}\left\vert
n\right\rangle \left\langle n\right\vert $, $\overline{n}$ being the initial
mean number of quanta for each mode. \ To solve numerically the master
equation (\ref{eq6}) we adopt the same coupling, $\lambda_{1x}=\lambda
_{2y}=\lambda$, $\lambda_{1y}=\lambda_{2x}=\lambda\tanh\left(  r\right)  $,
and the same decay rate for both excited electronic states, $\Gamma_{1}%
=\Gamma_{2}=\Gamma$, and the same decay rate for the vibrational modes,
$\gamma_{a}=\gamma_{b}=\gamma$. In Fig. 2, we have plotted the mean number of
quanta of mode $a$ against time (the evolution of mode $b$ is identical) for
different values of $\gamma$ and for $\overline{n}=2$, $\overline{n}_{th}%
=0.5$, $\Gamma=10$, and $r=1$ (which implies $\tanh\left(  r\right)
=\left\vert \frac{\lambda_{1y}}{\lambda_{1x}}\right\vert =\left\vert
\frac{\lambda_{2x}}{\lambda_{2y}}\right\vert \simeq0.76$). For an ideal
two-mode vacuum squeezed state, the mean number of quanta for each mode is
$\left\langle n_{a,b}\right\rangle =\frac{\tanh\left(  r\right)  ^{2}}%
{1-\tanh\left(  r\right)  ^{2}}\,$, which for $r=1$ gives $\left\langle
n_{a,b}\right\rangle \simeq1.4$. We can see in Fig. 2 that, for $\gamma
=0.001\lambda$ and $\gamma=0.01\lambda$, a mean number of quanta close to this
value is reached asymptotically, but this is not so for $\gamma=0.1\lambda$,
because of the competition between the engineered and thermal reservoirs.
Another parameter we have used to analyze the fidelity of the generated state
is the total variance $\left\langle \left(  \Delta\widehat{u}\right)
^{2}+\left(  \Delta\widehat{v}\right)  ^{2}\right\rangle $ of a pair of
EPR-like operators $\widehat{u}=\left\vert \varepsilon\right\vert \widehat
{x}_{a}+\frac{1}{\varepsilon}\widehat{x}_{b}$ and $\widehat{v}=\left\vert
\varepsilon\right\vert \widehat{p}_{a}-\frac{1}{\varepsilon}\widehat{p}_{b}$
\cite{duan-prl2000}, with $\widehat{x}_{\alpha}=\left(  \widehat{\alpha
}+\widehat{\alpha}^{\dagger}\right)  /\sqrt{2}$ and $\widehat{p}_{\alpha
}=-i\left(  \widehat{\alpha}-\widehat{\alpha}^{\dagger}\right)  /\sqrt{2}$,
$\alpha=a,b$. According to Ref. \cite{duan-prl2000}, a two-mode Gaussian state
is entangled if and only if $\left\langle \left(  \Delta\widehat{u}\right)
^{2}+\left(  \Delta\widehat{v}\right)  ^{2}\right\rangle <\varepsilon
^{2}+1/\varepsilon^{2}$. For $\varepsilon=$ $1$ and an ideal two-mode vacuum
squeezed state, the total variance is $\left\langle \left(  \Delta\widehat
{u}\right)  ^{2}+\left(  \Delta\widehat{v}\right)  ^{2}\right\rangle
=2e^{-2r}$, which for $r=1$ gives us $\left\langle \left(  \Delta\widehat
{u}\right)  ^{2}+\left(  \Delta\widehat{v}\right)  ^{2}\right\rangle
\simeq0.27$. As we can see in Fig. 3, this value is reached and approximately
reached for $\gamma=0.001\lambda$ and $\gamma=0.01\lambda$, respectively.
Again, for $\gamma=0.1\lambda$, the action of the thermal reservoir does not
allow the ideal generation of the two-mode entangled state. Instead of
applying unitary transformations to the density matrix $\rho$, which led to
Eq. (\ref{eq4}), and thus making it easy to find the steady state, we could
have proceeded by looking for an engineered Liouvillian for the engineered
reservoir, as in Ref. \cite{carvalho-prl2001}. For an atomic decay rate
$\Gamma$ much stronger than the effective coupling $\lambda$ and the decay
rate of the vibrational modes $\gamma$, the effective \textit{decay rate} for
the engineered reservoir is given by $\Gamma_{eng}=$ $4\lambda^{2}/\Gamma$. In
our numerical solution of the master equation (\ref{eq6}) we have used
$\Gamma=10\lambda$, which results in $\Gamma_{eng}=0.4\lambda$, that is of the
same order of magnitude as $\gamma=0.1\lambda$. Then, for this value of
$\gamma$, the influence on the generated TMVS state of the natural reservoir
is almost the same as that of the engineered reservoir. Hence, to minimize the
influence of the natural reservoirs, we must have $\Gamma_{eng}=$
$4\lambda^{2}/\Gamma\gg\gamma$. For example, for $\eta=0.1$ (Lamb-Dicke
limit), $\Omega_{i\alpha}\sim1$ MHz, and $\Gamma\sim1$ MHz, which can easily
be achieved with current technology, we have $\left\vert \lambda\right\vert
\sim0.1$ MHz and $\Gamma_{eng}=40$ KHz, which is much stronger than
$\gamma\sim2$ KHz, found in present-day experiments. (The chosen values above
also satisfy the requirements for the approximations employed to obtain the
effective Hamiltonian: for $\nu_{x}\sim\nu_{y}\sim30$ MHz $>>\left\vert
\eta\Omega\right\vert \sim0.1~$MHz and $\Gamma\sim1$ MHz.)

Summarizing, we have presented a simple scheme to prepare a two-mode vacuum
squeezed state for the 2D motion of a trapped ion via the generation of an
artificial reservoir. Our scheme is robust against the usual mechanism of
dissipation and could be implemented with the present-day technology and we
hope it could be employed to test experimentally the entanglement properties
of Gaussian states. The approximations employed in this work were validated by
numerical calculations, which showed complete agreement with the analytical
solutions. To prove the engineered two-mode state a tomographic method could
be employed that enables the Wigner function of the entangled state to be
reconstructed \cite{fwigner}.

We wish to acknowledge the support of the Brazilian agencies CNPq, FAPESP
(process No. 2005/04105-5), and Brazilian Millennium Institute for Quantum Information.

Figure Captions:

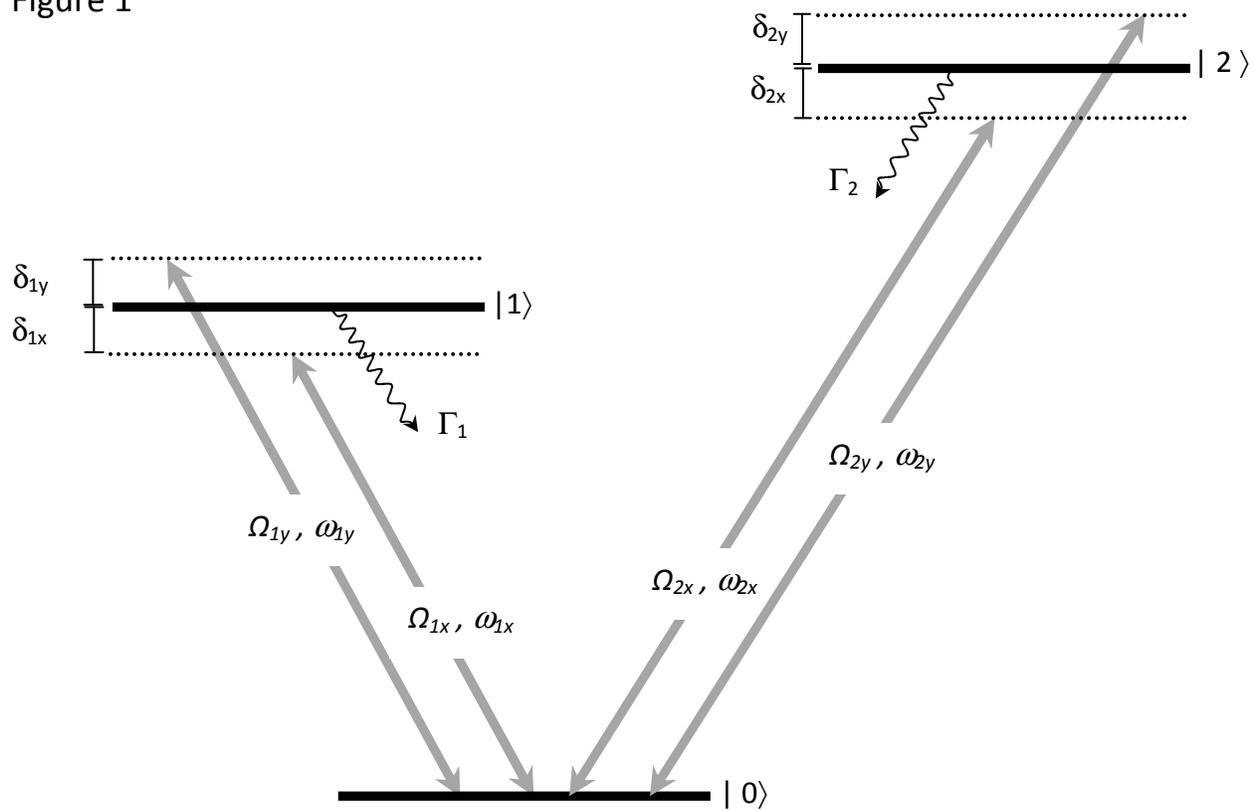
Fig. 1: Atomic levels of the trapped ion. The ground state $\left\vert
0\right\rangle $ is coupled to the excited states $\left\vert 1\right\rangle $
and $\left\vert 2\right\rangle $ through laser fields.

Fig. 2: The time evolution of the mean number of quanta, $\left\langle
a^{\dagger}a\right\rangle $, of the vibrational mode $x$ for $\Gamma
=10\lambda$, $\overline{n}_{th}=0.5$, $r=1$, and three values of the decay
rate of the vibrational modes: $\gamma=0.001\lambda$ (solid line),
$\gamma=0.01\lambda$ (dotted line), and $\gamma=0.1\lambda$ (dashed-dotted
line). The dashed line (straight line) represents the expected value.

Fig. 3: The time evolution of the total variance $\left\langle \left(
\Delta\widehat{u}\right)  ^{2}+\left(  \Delta\widehat{v}\right)
^{2}\right\rangle $ for the same parameters used in Fig. 2.

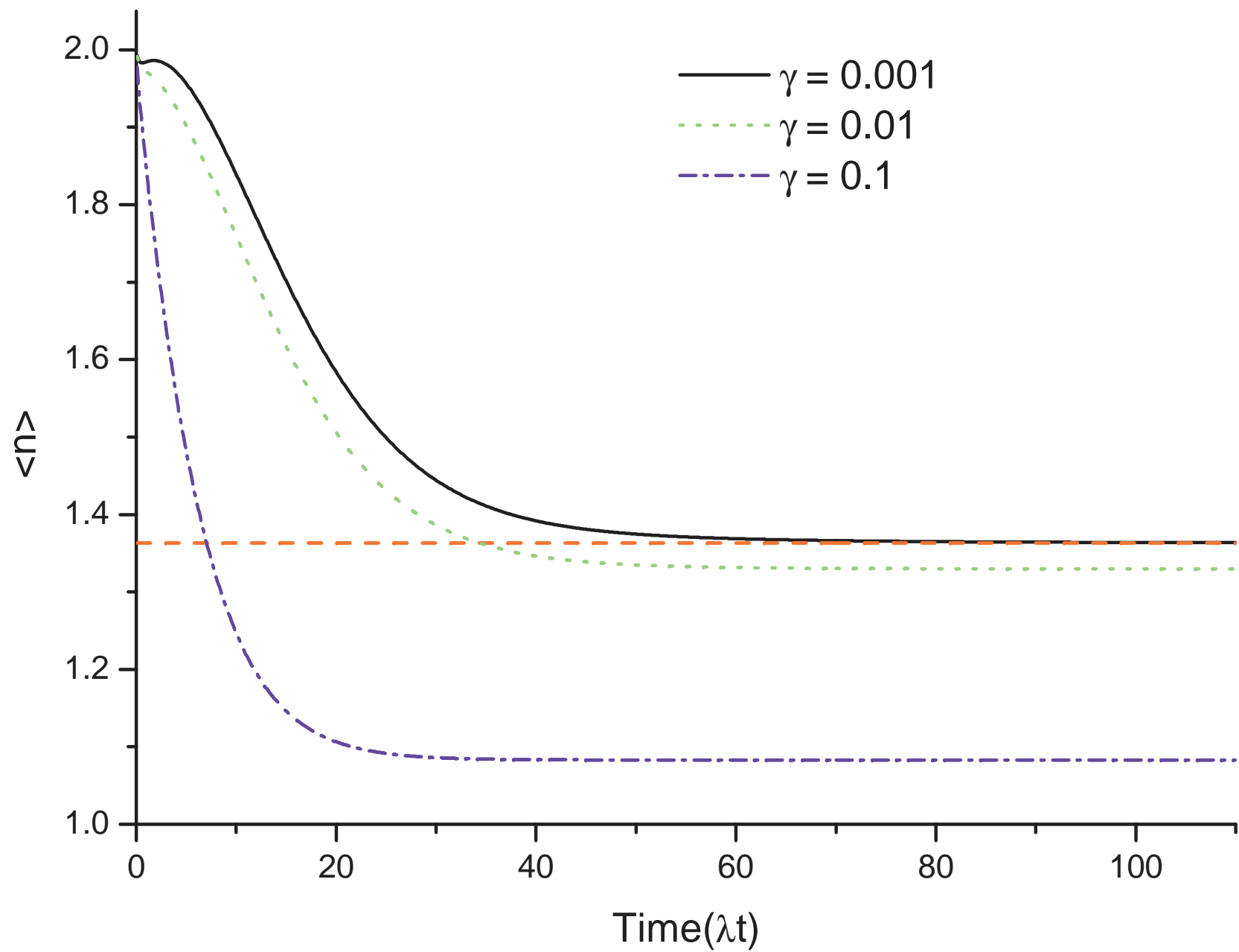
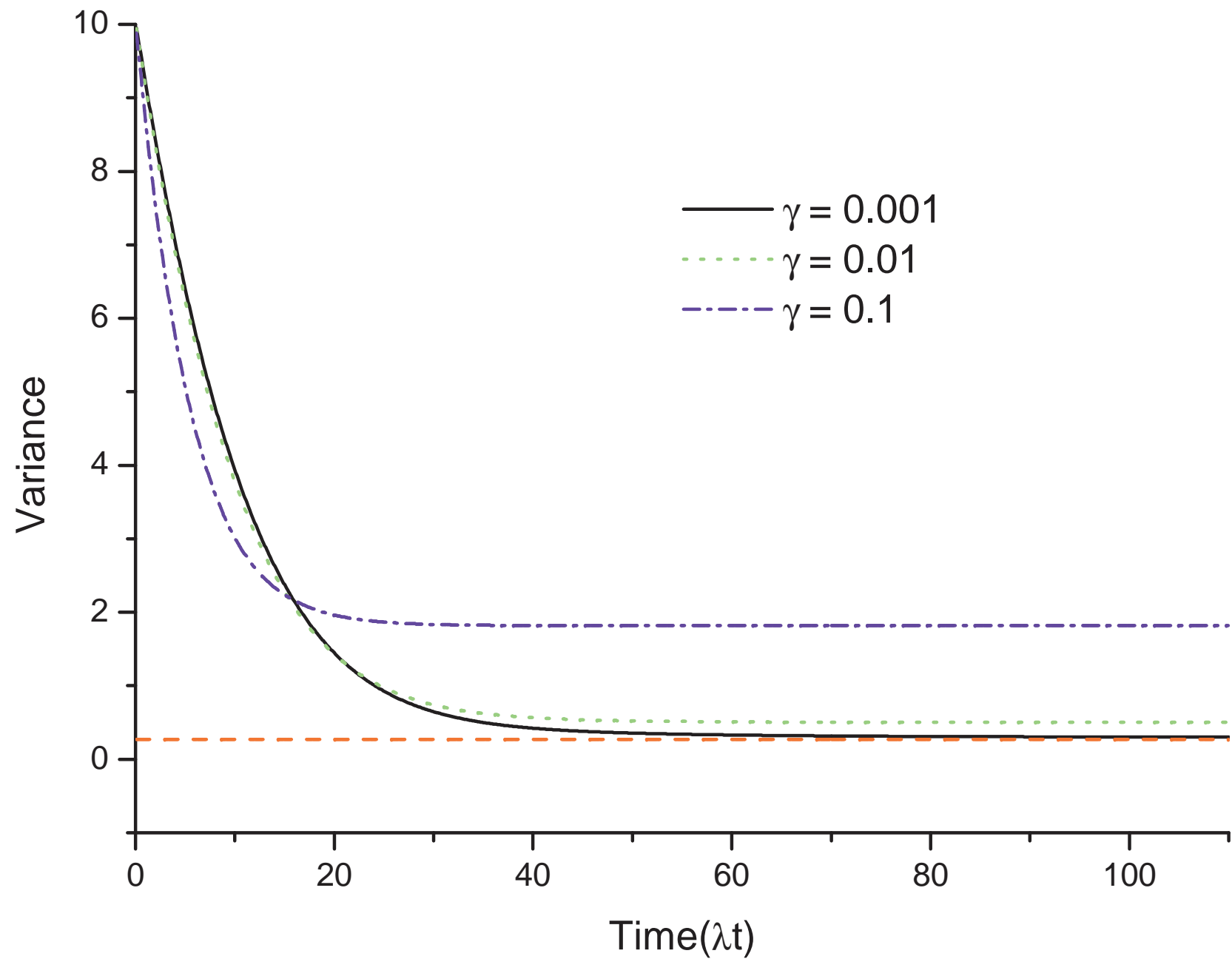
\end{document}